\documentclass{elsarticle}
\usepackage{graphicx}
\usepackage[colorlinks=true,urlcolor=blue]{hyperref}

\begin{document}



\title{Elimination of the effect of internal activity in LaCl$_{3}$:Ce scintillator}

\author{D. Chattopadhyay \corref{cor}}
\cortext[cor]{Corresponding author.}
\ead{dipayanchattopadhyay90@gmail.com}
\author{Sathi Sharma} 
\author{M. Saha Sarkar}

\address{Saha Institute of Nuclear Physics, HBNI, 1/AF Bidhannagar, Kolkata - 700 064, India}
\date{\today}

\begin{abstract}
The Lanthanum Halide scintillator detectors have been widely used for nuclear spectroscopy experiments because of their excellent energy and time resolutions. Despite having these advantages, the intrinsic $\alpha$ and $\beta$ contaminations in these scintillators pose a severe limitation 
in their usage in rare-event detections. In the present work, pulse shape discrimination (PSD)  with a fast digitizer has been shown to 
be an efficient method to separate the effect of $\alpha$ contamination from the spectrum. The shape of the $\beta$ spectrum has been generated with the 
help of Monte Carlo based simulation code, and its contribution has been eliminated from the spectrum. 
The reduction in the background events generated by both intrinsic $\beta$ and $\alpha$ activities has been demonstrated.
The present study will encourage the application of these detectors in low cross-section measurement experiments relevant to nuclear astrophysics. 
 
\end{abstract}

\maketitle

\section{\label{sec:level1}Introduction}
Commercially available La-halide detectors have been remarkably successful in the field of radiation measurement. Cerium-activated La-halide detector offers brilliant light output, excellent energy resolution, a fast response, excellent linearity ~\cite{rosza,vanLoef02,menge07,glodo05,quarati07,moses02,shah04} and a stable light output over a wide range of temperatures~\cite{moszynski06}. For Cerium-activated  LaCl$_{3}$ detector, the light output is  $\sim$50 photons/keV with energy resolution as low as 3.1$\%$ at 662 keV. It has fast response time with principal decay constant $\sim$ 20 ns,  and non-linearity below 7$\%$ from 60 to 1300 keV~\cite{loef01,shah03}. This combination of features makes La-halide detectors useful for low energy nuclear spectroscopy experiments~\cite{loef001}, time of flight measurements~\cite{Schaart10} and medical imaging purposes~\cite{witherspoon10}. La-halide detectors have been used for the lifetime measurements of unstable nuclei ~\cite{white07,marginean10}, and even for detecting the fusion neutrons~ \cite{cazzaniga13}. Furthermore, due to the high Z of Lanthanum 
(Z=57) and high density of the crystal (for, LaCl$_{3}$, $\rho$=3.85 g/cm$^{3}$), large volume La-halide  detectors can be useful for detection of high energy $\gamma$-rays (up to 20 MeV). However, the self-activity of the La-halide detector has been observed to be a major issue that reduces the detector sensitivity and interferes with the $\gamma$ rays of interest in nuclear physics experiments and complicates data analysis. 

Natural Lanthanum is composed of stable $^{139}$La with 99.91$\%$  abundance. The radioactive $^{138}$La with half-life 1.05$\times 10^{11}$ y contribute to the remaining 0.09 \% of  the abundance \cite{meija16}. Since the separation between the two isotopes is almost impossible, the contamination due to $^{138}$La is present in all La-halide based scintillators. $^{138}$La isotopes decay by electron capture (e$^{-}$) into $^{138}$Ba with 66.4 $\%$ probability and the remaining 33.6 $\%$ decay by $\beta^-$ - emission  into $^{138}$Ce. In both cases, $^{138}$La decays into an excited state of the corresponding daughter nucleus resulting in the emission of subsequent $\gamma$ rays of equivalent energy. For the electron capture process, 1436 keV $\gamma$-ray along with 32-38 keV characteristic X-rays of Ba is emitted. Whereas, in the second case, the emission of 789 keV $\gamma$-ray in coincidence with the $\beta$-continuum having endpoint energy of 255 keV is observed. 

Moreover, actinium is a chemical analogue to lanthanum, {\it i.e.}, they have very similar chemical properties. All the actinium isotopes are radioactive. The longest-lived actinium isotope is  $^{227}$Ac having a half-life of 21.772 y.  La-halide based scintillators mostly contain actinium impurities, which contribute to the $\alpha$ contaminations due to the decay of long-lived $^{227}$Ac isotopes. The inherent radioactivity  results in an intrinsic background of about 1-2 counts/cm$^3$s \cite{knoll}. As the $\alpha$/$\beta$ energy loss ratio is less than 1, which is typical of nearly all scintillators, the $\alpha$ peaks appear at lower energies than their actual energies when the scale is calibrated from $\gamma$-ray (or fast electron) energies \cite{knoll}. The self-activity background due to $\alpha$ decay from Ac-series nuclei appears at a range of 1.5-3 MeV in the energy spectrum. The contribution of $^{227}$Ac component from impurities has been significantly reduced with the improved techniques of growing the lanthanum halides. This inevitable background can be negligible if the count rate of actual events is sufficiently larger than self-activity background events. However, it is difficult to estimate the real event rate for a low count rate experiment. 

In general, for high energy $\gamma$-ray measurements, large volume La-halide detectors can
be useful, but the event rate of self-activity increases with crystal volume. For smaller crystals, the presence of the 1436 keV peak indicates that a significant fraction of the  X-rays escape without detection. However, for larger crystals, these X-rays are summed with 1436 keV $\gamma$s to show a dominant sum peak at about 1468 keV \cite{knoll}. Using Cerium bromide instead of La-halide, one can obtain higher radio-purity, but CeBr$_{3}$ has worse energy resolution compared to other La-halide scintillator \cite{meija16}. Therefore, research on self-activity elimination has significant importance for the La-halide detector to overcome the limits of their applicability in low energy nuclear astrophysics experiments where the event rate is low.   

Several studies have been done to study the internal activity of these detectors and find a way to eliminate them from the spectrum.  Hartwell and Gehrke ~\cite{hartwell05} identified the presence of $^{227}$Ac and its daughters as contamination of $\alpha$-emitting nuclides in LaCl$_{3}$:Ce. Milbrath {\it et al.} ~\cite{milbrath05} also confirmed the presence of $^{227}$Ac as the source of $\alpha$ contamination in LaCl$_{3}$:Ce through coincidence measurements. Szucks {\it et al.} ~\cite{szucs12} in their paper have also pointed out the demerits of the usability of LaBr$_{3}$ detector in low energy nuclear astrophysics experiments within the desired energy range because of the mixing of contamination due to internal activity with the real counts. Studies by Hoel {\it et al.} ~\cite{hoel05} and Crespi {\it et al.} ~\cite{crespi09} suggested the existence of pulse shape differences between $\alpha$ and $\gamma$ pulses. Hoel {\it et al.} ~\cite{hoel05} have also observed that for LaCl$_{3}$:Ce, the difference in pulse shapes is modest but for LaBr$_{3}$ the difference is too small to be useful. Crespi {\it et al.} ~\cite{crespi09} by using the charge comparison method with a fast digitizer achieved the suppression of intrinsic $\alpha$ background. The radioactive decays of $^{138}$La in LaBr$_{3}$:Ce and the shape of the internal $\beta$ spectrum have been measured by Quarati {\it et al.} ~\cite{quarati12}. Recently, the shape of intrinsic alpha pulse height spectra in lanthanum halide
scintillators have been studied in detail by Wolszczak and  Dorenbos \cite{2017}. Despite these efforts, the overall quantitative understanding of the effect of internal activity in the spectrum of the La-halide detector is still limited now.  Lanthanum halide scintillators, namely LaCl$_{3}$:Ce and LaBr$_{3}$:Ce, show very similar trends in terms of internal activity and, in principle, the results obtained using one of them are directly applicable to the other. In the present work, the internal radioactivity of a LaCl$_{3}$:Ce detector has been studied. 
  
This study aims to provide a deeper understanding of the internal $\alpha$ and $\beta$ activities in LaCl$_{3}$ and find a way to eliminate them from the spectrum so that the detector can be useful for the low energy capture cross-section measurements relevant to nucleosynthesis.   
In this paper, we present the pulse shape discrimination (PSD) technique to eliminate the effect of internal $\alpha$ contamination from the spectrum in LaCl$_{3}$ detector. The PSD parameter is optimized for better separation between $\alpha$ and $\gamma$ pulses. Besides, the Monte Carlo based simulation code using GEANT4 toolkit has been developed to eliminate the effect of $\beta$ activity from the spectrum. 

The paper is organized as follows. In section 2, the details of the experimental setup is discussed. The different techniques used for the elimination of the internal activity are discussed in section 3. In section 4, results obtained from the different methods have been analyzed and discussed. Finally, the summary and conclusions are addressed in section 5.  	
	
\section{\label{sec:level2}Experimental Details}	 
\begin{figure}  
\begin{center}
\includegraphics[scale=0.5]{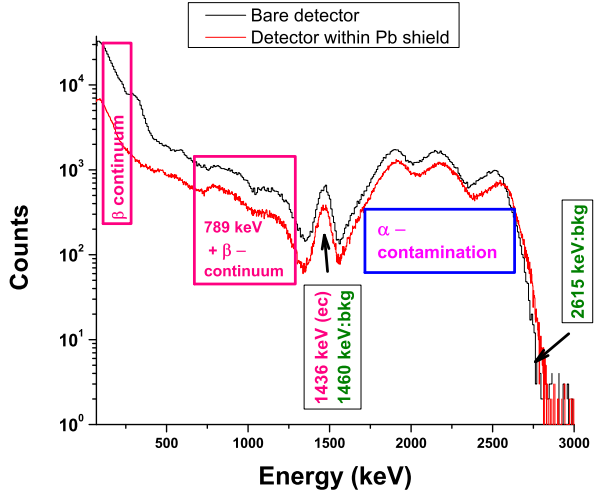}\vspace{-4cm}

\caption{\label{fig:bkgpb}(a) Normal room background spectrum acquired with and without lead shielding. 
The room background $\gamma$-ray  peaks,  1460 keV from $^{40}K$, and 2615 keV (hidden under the alpha contamination peaks) from $^{232}Th$ decay series are shown. The intrinsic background components arising from the decay of $^{138}La$: $\beta$-continuum, the summing bump of 789 keV ($\beta^-$ decay) with the $\beta$ continuum, 1436 keV $\gamma$- peak (electron capture) are shown. The peaks arising from $\alpha$-contamination in the detector from $^{227}Ac$ impurity are also indicated in the figure.  }
\end{center}
\end{figure}

In this work, an $1^"\times 1^"$ cylindrical LaCl$_{3}$:Ce scintillator crystal from Saint-Gobain coupled with a Hamamatsu photo-multiplier tube (PMT) was used. The PMT has been biased to -1800 V. The anode signal was processed by a CAEN digitizer DT5730 (sampling rate 500 MHz). The data were acquired in list mode, {\it i.e.},  event by event mode, and further analyzed in off-line by LAMPS ~\cite{chatterjee} data analysis software. Low activity $^{60}$Co, $^{152}$Eu, $^{133}$Ba, $^{137}$Cs, $^{241}$Am and $^{207}$Bi sources were used for energy calibration purposes. To minimize the contribution of 
room background $\gamma$-radiation in the spectrum,  the detector is kept inside a lead shielding. The spectra with and without Pb shielding are shown in Fig. \ref{fig:bkgpb}. 

\section{\label{sec:level3}Techniques used for elimination of the internal activity}
\subsection{\label{sec:level3a}Pulse shape discrimination }
When an energetic particle or photon is incident on a scintillator crystal, it deposits its energy in the medium. The scintillator gets excited and emits scintillation radiation. The energy loss per unit thickness (dE/dx) inside the scintillator material varies for different incoming particles. Hence, the luminescence rates for different particles result in pulse shape differences. Two exponentials with two-time components \cite{decayt} usually characterize the time evolution of the intensity of the scintillation light. One is a fast or prompt component, and the other is a slow or delayed component. The time evolution of the number of emitted scintillation photons N from a single scintillation event can be described by a linear superposition of two components given by 

\begin{equation}
N=\frac{A}{\tau_{f}} e^{-\frac{t}{\tau_{f}}}+ \frac {B}{\tau_{s}} e^{-\frac{t}{\tau_{s}}}
\end{equation}

where, $\tau_{f}$ and $\tau_{s}$ are the fast and slow decay times. The majority of the scintillation intensity is contained in the prompt component.
However,  the long-lived or the slow component has an important implication as the fraction in it often depends on the nature of the incident particle. This dependence can be utilized to differentiate between particles of different kinds that deposit the same energy in the detector 
\cite{knoll}. This technique is well known as pulse shape discrimination (PSD). It is widely applied to eliminate unwanted events in a mixed radiation field. The light emission intensity of a LaBr$_3$:(Ce) scintillator is accurately modeled with a single fast component exponential decay. However, that of a LaCl$_3$:(Ce) crystal is represented well by a two-component exponential decay with  $t_f$ $\simeq$ 26 ns
and $t_s \simeq$ 550 ns at 20$^o$C. Moreover, the ratio of B/A is approximately
0.25 for $\gamma$-rays at 20$^o$C~\cite{Mcfee16}.

\begin{figure} [h]
\includegraphics[width=.8\textwidth]{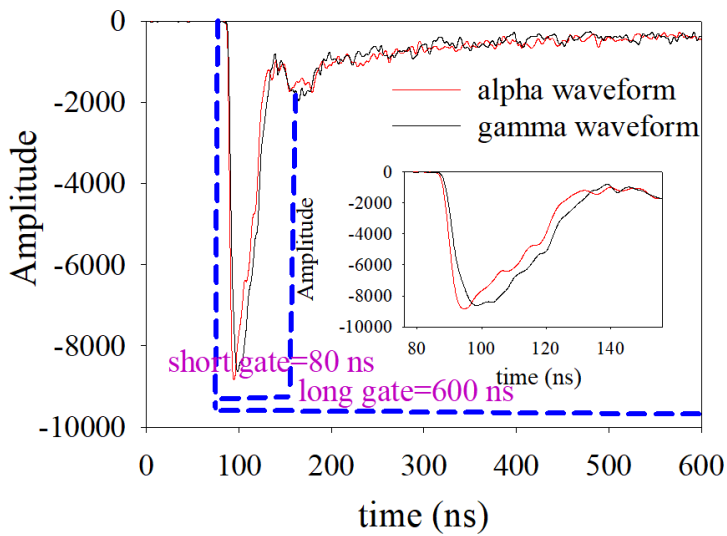}
\caption{\label{fig:gamma_alpha} Typical waveforms of $\alpha$ and $\gamma$ obtained from LaCl$_{3}$ detector having the same total charge deposited. The short and long gates are 
shown in the figure. The inset highlights the difference between an $\alpha$ and a $\gamma$ pulse. }
\end{figure}
In pulse shape discrimination with scintillators, the most frequently used technique is the charge integration method, which determines the amount of delayed light output with respect to the total light output for each event to identify the type of the corresponding ionizing radiation. DPP-PSD firmware 
provided with the CAEN 5730 digitizer \cite{CAEN} is based on this method. The PSD parameter is then extracted in event-by-event mode using the FPGA of the digitizer as, 

\begin{equation}
PSD=\frac{(Q_{l} - Q_{s})}{Q_{l}}
\end{equation}

where, $Q_{l}$ and $Q_{s}$ are the integrated charge within the long gate and short gate,  respectively, as shown in Fig.~\ref{fig:gamma_alpha}. The PSD is the ratio of charge deposited at the tail part of the pulse ($Q_{l}$ - $Q_{s}$) to the total charge deposited in the full pulse ($Q_{l}$). The short window (80 ns in the present case) is to be chosen in such a way that the ratio of the tail to the total pulse would be most effective for particle identification. For example, $\alpha$ and $\gamma$ have different interaction mechanisms with the detector material; hence they have different responses in the detectors (shown in the inset of Fig.~\ref{fig:gamma_alpha}). For the same total charge deposited, $\alpha$ pulses are narrower and sharper compared to $\gamma$ pulses. The choice of short and long gates  is shown in Fig.~\ref{fig:gamma_alpha}. With this choice of gates, the PSD parameter is larger for $\alpha$s than for $\gamma$s. 

\begin{figure} [h]
\begin{center}
\includegraphics[scale=0.70]{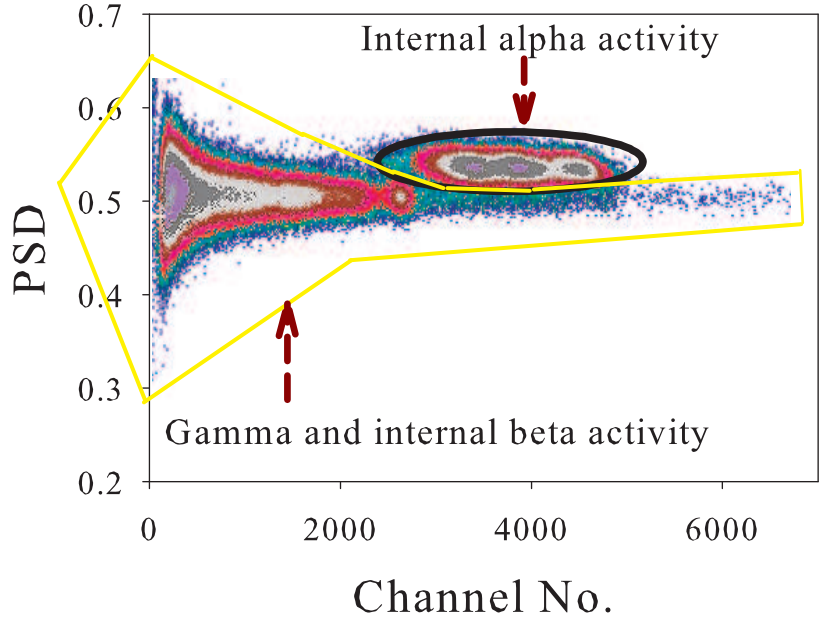}
\caption{\label{psdbkg} 2D plot of PSD vs Channel No. (Energy) for  normal room background for the Pb-shielded  detector.}
\end{center}
\end{figure}

This technique has been used to eliminate the contributions of internal activity due to $\alpha$ contamination from the spectrum. A two-dimensional plot of PSD vs Channel no. ($Q_{l}$, {\it i.e.}, energy) shown in Fig.~\ref{psdbkg} distinguishes the $\alpha$ contamination events from the $\gamma$'s and 
$\beta$'s. 

\subsection{\label{sec:level3b}Simulation based on Monte Carlo method }
LaCl$_{3}$:Ce detector has an internal $\beta$ activity that can not be eliminated using the pulse shape discrimination as the shape of the pulses looks very similar for both $\gamma$ and $\beta$. For the elimination of the effect of internal $\beta$ activity from the $\gamma$ spectrum of LaCl$_{3}$:Ce, a Monte Carlo based simulation code has been developed using the GEANT4 toolkit~\cite{agostinelli16}. In the present work, the GEANT4 toolkit version 4.10.0 has been implemented.  

The three main classes in our code are detector construction, physics list, and primary generator action. In the detector construction class, the geometry and materialistic information of the detector and its encasing  is defined as provided by the manufacturer. The geometry consists of a small 
$1^" \times  1^"$ cylindrical LaCl$_{3}$ detector surrounded by MgO reflector, and optically coupled to a Bi-alkali Photo-cathode through a Quartz PMT Window. In the physics list, the necessary physics processes, such as, G4EmStandardPhysics, G4DecayPhysics, and G4RadioactiveDecayPhysics are included. The G4EmStandardPhysics class contains the three standard electromagnetic processes like Compton scattering, Photoelectric process, and Pair production. The G4DecayPhysics constructor handles the decay channels for all unstable particles defined in the physics list. The same process is assigned to all unstable particles. The G4RadioactiveDecay\-Physics contains the basic features of the radioactive decay of nuclei. In, primary generator action class, the general particle source (GPS) module has been used as a particle generator to create different shapes with a specific position, angle, and energy distribution, etc. For the simulation of internal $\beta$ activity, a cylindrical source of  same size as the detector has been used.   

The simulations were carried out for a large number of events (10$^5$) to reduce the statistical uncertainty. The response function of the internal $\beta$ activities was generated using the simulation code. The detailed procedure  is described below.

No. of $^{138}$La atoms (N$_{0}$) inside the LaCl$_{3}$ detector has been estimated from the  following relation.
\begin{equation}
N_{0}=\frac{N_{A}.\rho.(\pi r^{2}h).(f)}{A}
\end{equation}    

where N$_{A}$ is the Avogadro number, $\rho$ is the density of the detector, $r$ is the radius, and $h$ is the height of the cylindrical detector, 
$f$ is the fractional abundance of $^{138}$La in  mass A of LaCl$_{3}$.

The number of $^{138}$La atoms (N) remaining after time $\tau$ can be calculated as:
\begin{equation}
N=N_{0}exp(-\lambda.\tau)
\end{equation}  

Then the  number of $^{138}$La atoms which have decayed in this time would be:
\begin{equation}
N_{0}-N=N_{0}(1-exp(-\lambda.\tau))
\end{equation}

It is known that in 66.4$\%$ times $^{138}$La decays to $^{138}$Ba via an electron capture with the emission of $\sim$ 1436 keV $\gamma$ line. Hence in the elapsed time $\tau$ no. of emitted $\gamma$s (N$_{1436}$) having an energy of $\sim$ 1436 keV can be estimated as:
\begin{equation}
N_{1436}=0.664*N_{0}(1-exp(-\lambda.\tau))
\end{equation}

The simulated spectra (Fig. \ref{fig:simulated_activity}) have been normalized to reproduce the N$_{1436}$ counts observed in the experimental spectra to get the shape of the 
internal $\beta$ activity spectrum.  The contribution of the 1460 keV room- background $\gamma$ - peak in N$_{1436}$ has been minimized as far as possible.
   
The $\gamma$ energy spectra of LaCl$_{3}$ detector for  chosen energies have been generated by assuming a simplified expression of the peak shapes.
\begin{equation}
\sigma(E)=a+b\sqrt{E}
\end{equation}

where, $\sigma(E)$ is the standard deviation of the peak shape of the $\gamma$ peak of energy E in the $\gamma$ spectrum. 
The standard deviation is obtained from the full width at half maximum (FWHM) of a $\gamma$ peak.  The value of the parameters $a$ and $b$ have been estimated by  fitting the variation of FWHM of $\gamma$ peaks as a function of peak energies obtained from spectra of different radioactive  sources 
($^{60}$Co, $^{137}$Cs and $^{207}$Bi).

\section{\label{sec:level4}Results and Discussions}
\subsection{\label{sec:level4a}The background spectra}
Fig.~\ref{fig:bkgpb}  shows the background spectrum obtained from LaCl$_{3}$ detector acquired with and without lead shielding. Due to the inherent radioactivity in the detector, the background spectrum is dominated by components of $^{138}$La and $^{227}$Ac decay.  In the electron capture decay
mode of $^{138}$La, a  $\gamma$-ray (1436 keV) and X-ray (32 keV-K$_{\alpha}$ and 38 keV-K$_{\beta}$) from $^{138}$Ba are emitted. The 1436 keV $\gamma$ and the 1460 keV $\gamma$ from the $^{40}$K present in the room background can not be resolved. For relatively larger detectors the correlated $\gamma$-ray and K X-rays(32-38 keV) of $^{138}$Ba give rise to a sum-peak  at $\sim$ 1470 keV ~\cite{menge07,roberts}.

Following the $\beta^-$ decay, a continuum $\beta^-$ spectrum till 255 keV followed  a by 789 keV $\gamma$-ray from $^{138}$Ce are emitted.  The correlation  between $\beta$ particles (till 255 keV) and $\gamma$-ray (789 keV) of $^{138}$Ce generates a summed structure of  $\beta$ continuum
with $\gamma$-ray. The continuum structure  starts at 789 keV and  spreads to high energy at 1044 keV ~\cite{menge07,roberts}. The remaining 
$\beta$ continuum is observed at the  low energy side from 0 keV to 255 keV. In between 255 keV to 750 keV  the Compton continuum from the 789 keV and 1436 keV $\gamma$-rays are observed. Above 1700 keV, the presence of  $\alpha$ contaminant peaks are observed. These peaks originate from 
the $\alpha$'s emitted from the decay of $^{227}$Ac contamination. The $\alpha$'s  produce  a broad response with several peaks from roughly 1.7-2.3 MeVee (MeV electron-equivalent) ~\cite{roberts}.  Thus the background spectrum obtained from LaCl$_{3}$ detector is quite complicated and background components due to natural radioactivity are mostly hidden under the intrinsic background. 

\begin{figure}[h]
\begin{center}
\includegraphics[width=1.5\textwidth]{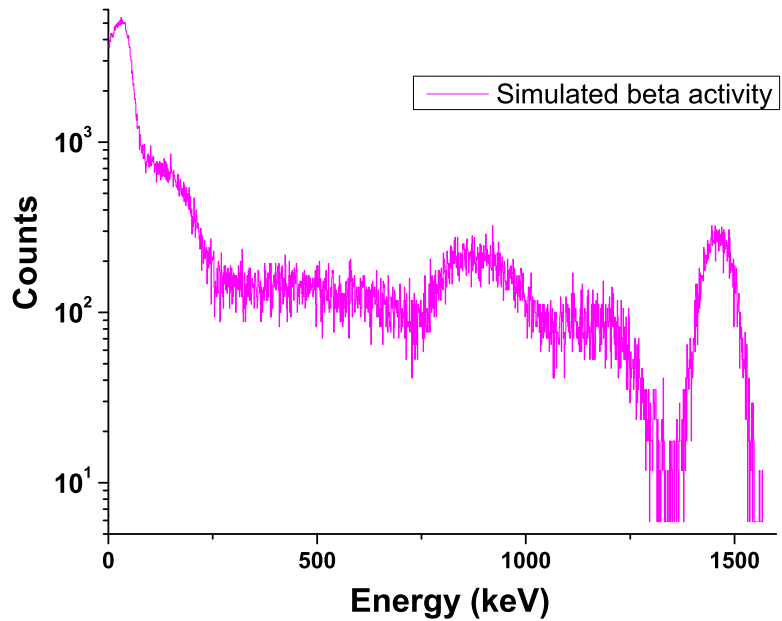}
\end{center}
\vspace{-5cm}\caption{\label{fig:simulated_activity} Shape of the $\beta$ spectrum obtained from the simulation.}
\end{figure}

\subsection{\label{sec:level4b}The $^{207}$Bi spectra: Elimination of the effect of $\alpha$ contamination from the spectrum}
\begin{figure} [h]
\begin{center}
\includegraphics[scale=0.75]{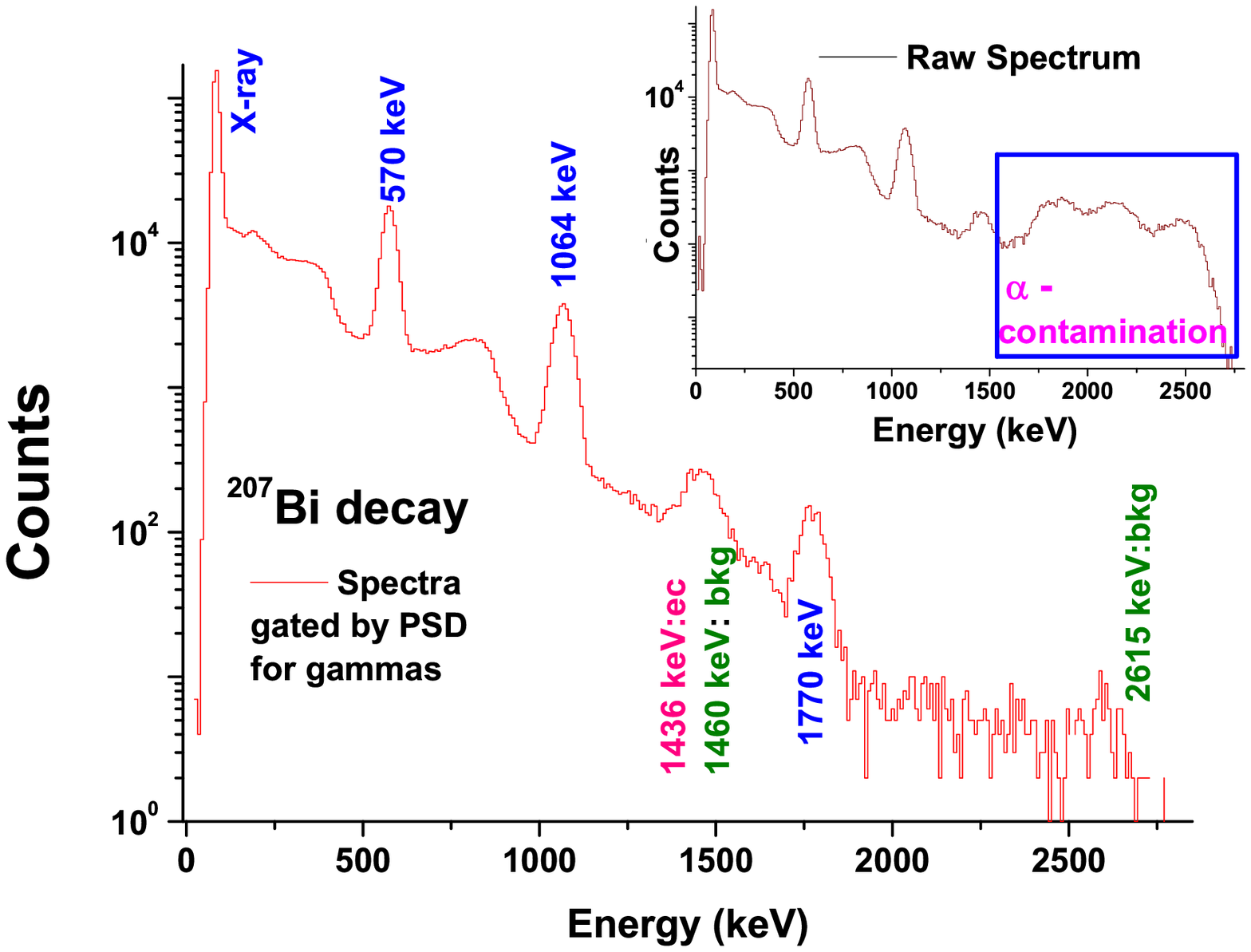}
\end{center}
\vspace{-6cm}
\caption{\label{fig:en} Raw (inset) and $\alpha$-activity subtracted spectra for  $^{207}$Bi decay.}
\end{figure}
The energy spectra using $^{207}$Bi source using a Pb-shielded detector acquired with DT5730 digitizer has been shown in Fig.~\ref{fig:en}. It is known that $^{207}$Bi source emits three $\gamma$s at 570 keV, 1064 keV, and 1770 keV. From Fig.~\ref{fig:en} the intrinsic activity of LaCl$_{3}$:Ce is found to be dominant in the spectrum.  Although 570 and 1064 keV peaks of $^{207}$Bi are seen, 1770 keV peak is invisible. The 1770 keV peak is hidden under the $\alpha$ peaks arising from the $\alpha$ activity in the detector. Therefore, proper elimination of the effect of internal activity from the spectrum is needed.

The effect of the contamination due to $\alpha$ activity is eliminated from the spectrum by using pulse shape discrimination. The PSD has been calculated event-by-event from the formula mentioned above. Fig.~\ref{psdbkg} shows a typical two-dimensional plot of PSD vs Channel No. (integral of Long Gate). The plot shows that one can separate the contributions for the $\alpha$ events from the $\gamma$ events. The $\alpha$ subtracted spectrum has been generated by plotting only those events which have PSDs corresponding to $\gamma$ events. The spectra thus generated is shown in Fig.~\ref{fig:en}. After the invoking PSD gate, the 1770 keV peak of $^{207}$Bi source is observed clearly in the $\alpha$ contamination subtracted spectrum.

\subsection{\label{sec:level4c}Estimation of reduction factor of intrinsic background in LaCl$_{3}$}
 The detector is kept inside a 2cm  lead shielding to reduce background $\gamma$ radiation. In this way, the contributions from the room background have been reduced by a factor of $\sim$ 11 $\%$ (Fig. \ref{fig:bkgpb}). However, the spectra are not free from intrinsic background. Elimination of internal $\alpha$ and $\beta$ contamination from the spectrum is required to reduce the background further. 
\begin{figure}  
\begin{center}
\includegraphics[scale=0.55]{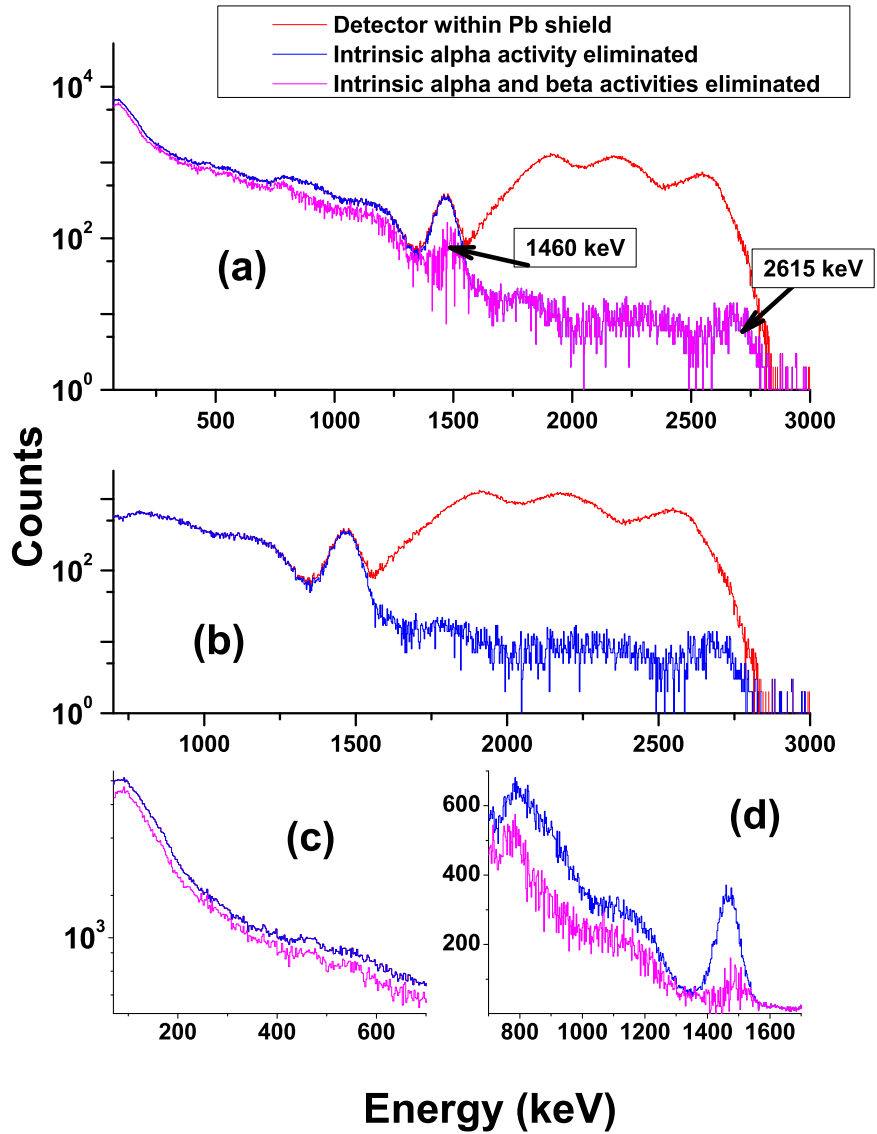}

\caption{\label{fig:bkg}(a) Normal room background spectrum acquired with  lead shielding compared with  that after elimination of $\alpha$ contamination, and also with that after elimination of both $\alpha$ and $\beta$ activities. The room background $\gamma$-ray  peaks,  1460 keV from $^{40}K$, and 2615 keV from $^{232}Th$ decay series are clearly seen after the elimination of the intrinsic background.  (b) The effect of removal of 
$\alpha$ activity from the spectrum, (c) and (d) results of elimination of $\beta$ activity from an $\alpha$ activity eliminated spectrum at (c) low energies ($<$700 keV) and (d) in the range from 700 -1700 keV. 
 }
\end{center}
\end{figure}

Pulse shape discrimination method has been adopted to eliminate the effect of $\alpha$ activity from the spectrum. 
Using the two-dimensional plot of PSD vs Channel No. (integral of Long gate) for background spectra the contribution of $\alpha$ contamination is eliminated, as shown in Fig.~\ref{fig:bkg} (b). 

The $\alpha$ subtracted spectrum still contains  the effect of $\beta$ contamination. The contribution of $\beta$ activities and the shape of the spectrum due to $\beta$ contamination alone has been simulated using GEANT4 based simulation code. In Fig.~\ref{fig:simulated_activity}, the shape of the spectrum due to the $\beta$ contamination is shown.   Finally, the total intrinsic activity, {\it i.e.}, the sum of the activities due to $\alpha$ and $\beta$ decays, is subtracted from the raw spectrum. The reduction factor (Fig.~\ref{fig:bkg}) to estimate the suppression in room background, as well as the intrinsic background, is calculated.  The time normalized background events in a Pb-shielded detector after elimination of intrinsic activities has been reduced by $\sim$ 35$\%$ compared to a bare detector.

\section{\label{sec:level5}Summary and conclusions}
In the present work, intrinsic activity due to $\alpha$-decay events in the LaCl$_{3}$:Ce scintillator has been eliminated by the pulse shape discrimination method using a fast digitizer. The shape of the $\beta$ continuum has been simulated using the Monte Carlo based simulation code via GEANT4. The simulated spectra after proper normalization has been subtracted from the raw spectra for the elimination of the effect of $\beta$ activity. The applicability  of the La-halide detectors for the low cross-section measurement experiments in nuclear astrophysics has been improved by reduction of background events by 35\%.  The discrimination method presented in this paper would be especially useful for LaCl$_3$:Ce detectors, which have a slow part in the scintillation pulse useful for pulse shape discrimination.  





\section{Acknowledgements}
We thank Prof. C. C. Dey of Saha Institute of Nuclear Physics, Kolkata for providing us the detector during the experiment and Mr. S. Karan of Saha Institute of Nuclear Physics, Kolkata for the technical help.


\end{document}